\documentclass[apl,amsmath,amssymb,amscd,groupedaddress,superscriptaddress,reprint,]{revtex4-1}

\usepackage[pdftex]{graphicx}
\usepackage{hyperref}
\usepackage{float}
\usepackage{color,soul}

\begin{document}

\title{GaN directional couplers for integrated quantum photonics} 
\author {Yanfeng Zhang}
\altaffiliation {These authors contributed equally to this work.}
\affiliation {Institute of Photonics, SUPA, University of Strathclyde, Glasgow G4 0NW, UK}
\author {Loyd McKnight}
\altaffiliation {These authors contributed equally to this work.}
\affiliation {Institute of Photonics, SUPA, University of Strathclyde, Glasgow G4 0NW, UK}
\author {Erman Engin}
\altaffiliation {These authors contributed equally to this work.}
\affiliation {Centre for Quantum Photonics, H. H. Wills Physics Laboratory $\&$  Department of Electrical and Electronic Engineering, University of Bristol, Merchant Venturers Building, Woodland Road, Bristol, BS8 1UB, UK}
\author {Ian M. Watson}
\affiliation {Institute of Photonics, SUPA, University of Strathclyde, Glasgow G4 0NW, UK}
\author {Martin J. Cryan}
\affiliation {Centre for Quantum Photonics, H. H. Wills Physics Laboratory $\&$  Department of Electrical and Electronic Engineering, University of Bristol, Merchant Venturers Building, Woodland Road, Bristol, BS8 1UB, UK}
\author {Erdan Gu}
\email[Author to whom correspondence should be addressed. Electronic mail: ]{erdan.gu@strath.ac.uk}
\affiliation {Institute of Photonics, SUPA, University of Strathclyde, Glasgow G4 0NW, UK}
\author {Mark G. Thompson}
\affiliation {Centre for Quantum Photonics, H. H. Wills Physics Laboratory $\&$ Department of Electrical and Electronic Engineering, University of Bristol, Merchant Venturers Building, Woodland Road, Bristol, BS8 1UB, UK}
\author {Stephane Calvez}
\affiliation {Institute of Photonics, SUPA, University of Strathclyde, Glasgow G4 0NW, UK}
\author {Jeremy L. O'Brien}
\affiliation {Centre for Quantum Photonics, H. H. Wills Physics Laboratory $\&$  Department of Electrical and Electronic Engineering, University of Bristol, Merchant Venturers Building, Woodland Road, Bristol, BS8 1UB, UK}
\author {Martin D. Dawson}
\affiliation {Institute of Photonics, SUPA, University of Strathclyde, Glasgow G4 0NW, UK}
\date{\today}

\begin{abstract}Large cross-section GaN waveguides are proposed as a suitable architecture to achieve integrated quantum photonic circuits. Directional couplers with this geometry have been designed with aid of the beam propagation method and fabricated using inductively coupled plasma etching. Scanning electron microscopy inspection shows high quality facets for end coupling and a well defined gap between rib pairs in the coupling region. Optical characterization at 800 nm shows single-mode operation and coupling-length-dependent splitting ratios. Two photon interference of degenerate photon pairs has been observed in the directional coupler by measurement of the Hong-Ou-Mandel dip with 96$\%$ visibility.
\end{abstract}
\maketitle 

Directional couplers (DCs) are one of the key components for integrated optics, and are a fundamental building block in emerging topics such as integrated quantum photonics circuits\cite{thompson2011integrated}. To date, quantum photonic circuits have been demonstrated in low-index-contrast waveguide materials including silica-on-silicon and glasses \cite{politi08,Marshall09,Sansoni10}. However, the choice of technological platform to realize particular embodiments of quantum information processing is still wide open. Fast path and polarisation manipulation have been demonstrated in lithium niobate waveguide devices\cite{bonneau2011fast}. Here we propose and demonstrate GaN-based structures as being attractive for such applications. GaN has a relatively high refractive index which enables the fabrication of compact guided wave circuits. It also benefits from mature, versatile and scalable epitaxial technology on a variety of substrates including silicon. Additionally, its transparency in the visible and near-infrared regions makes it attractive for use in combination with efficient Si-based single-photon detectors. Finally, it offers the prospects for on-chip integration with single photon sources, either fabricated in GaN itself (such as spontaneous parametric downconverters exploiting the nonlinear properties of GaN) or in the form of the diamond nitrogen-vacancy center, taking advantage of the close refractive index match between diamond and GaN. In this Letter, we report the design, fabrication and characterization, including degenerate photon-pair interference, of GaN-on-sapphire DCs as building blocks towards a platform for integrated photonic quantum circuits. 

\begin{figure}[t]
\includegraphics[width=85mm]{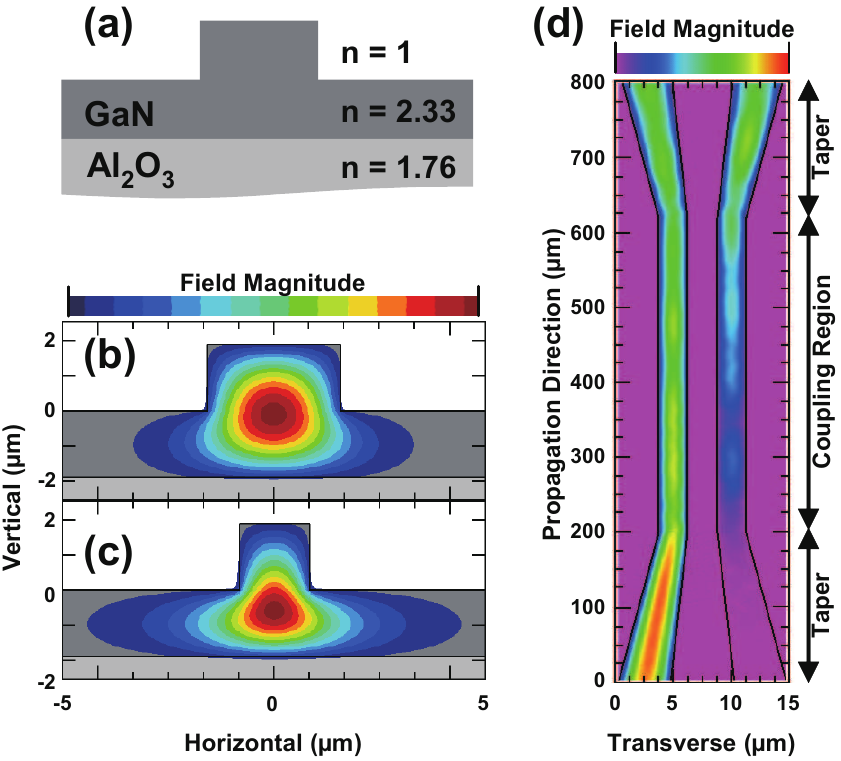}
\caption{\label{Simulation}(Color online) GaN DCs BPM simulations at a wavelength of 800nm. (a) Schematic of rib waveguide design using GaN on sapphire with air cladding. The ordinary index of GaN is used due to the restriction to TE operation\cite{Sanford03}. (b) TE field profile of guided mode at device input. (c) TE field profile of guided mode in coupling region. (d) Plan view BPM simulation of GaN DC design with a 2-$\mu$m gap, cross-sections vary from (b) to (c) in the coupler region.}
\end{figure}

\begin{figure}[t]
\includegraphics[width=85mm]{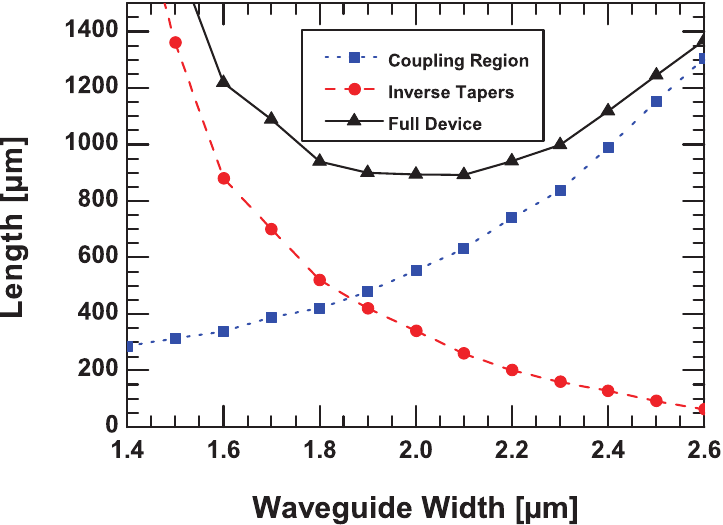}
\caption{\label{DeviceLength}(Color online) Device length optimization when changing rib width in the coupling region. It shows the minimum device length required to connect a 3.8-$\mu$m wide rib waveguide to the coupling section with a rib width between 1.4 and 2.6 $\mu$m for a 0.05 dB taper loss. }
\end{figure}

The design of the GaN-on-sapphire DCs (Fig. \ref{Simulation}) used in this study was specified to comply with a fabrication procedure involving standard optical contact lithography and a single etch step\cite{hui03,dahal09}. Structures were designed for samples of 3.8-$\mu$m thick \itshape c\upshape -plane GaN was grown on sapphire by metal-organic chemical vapour deposition. Designs were based on the widely-used large cross-section rib waveguide single-mode criteria \cite{Marcatili74,Soref91,Zhang11} that include near-circular profiles (numerically evaluated to be 95.8$\%$ with a circularly symmetric Gaussian beam). [Fig. \ref{Simulation}(b)] for efficient input and output end-coupling. All designs and simulations consider a free-space wavelength of 800 nm and TE (ordinary ray) operation. The key design feature is the use of optimized waveguide cross-sections and inverse tapers \cite{Moerman97} in order to enhance coupling efficiency and minimize device footprint. More specifically, as illustrated in Fig. \ref{DeviceLength} and supported by extensive simulations using the beam propagation method (BPM), for a fixed separation between the coupling waveguides (taken to be 2 $\mu$m) and a fixed rib etch depth (1.9 $\mu$m), the narrower the rib width, the shorter the coupling length. This originates from the fact that each mode spreads further horizontally into the underlying slab with reducing rib width [Fig. \ref{Simulation}(b) and (c)] which, in turn, leads to an increased mode overlap and coupling strength. However, when using inverse tapers to connect the input/output straight sections to the coupling region, the larger the rib width difference, the longer the taper needs to be to limit the amount of induced radiation loss. The optimization of the overall device footprint consists in selecting the width of the waveguide at the coupling section which is a trade-off between a short coupler and short inverse tapers. The full device length is calculated by BPM and is the summation of the constituent inverse taper and coupling region length. Fig. \ref{DeviceLength} shows that for device parameters consistent with the experimental ones, a waveguide width of 2 $\mu$m minimizes the device length at 900 $\mu$m. This offers \textit{a greater than four times reduction} in device length compared with a similar DC without inverse tapers and a 2-$\mu$m separation.

\begin{figure}
\includegraphics[width=85mm]{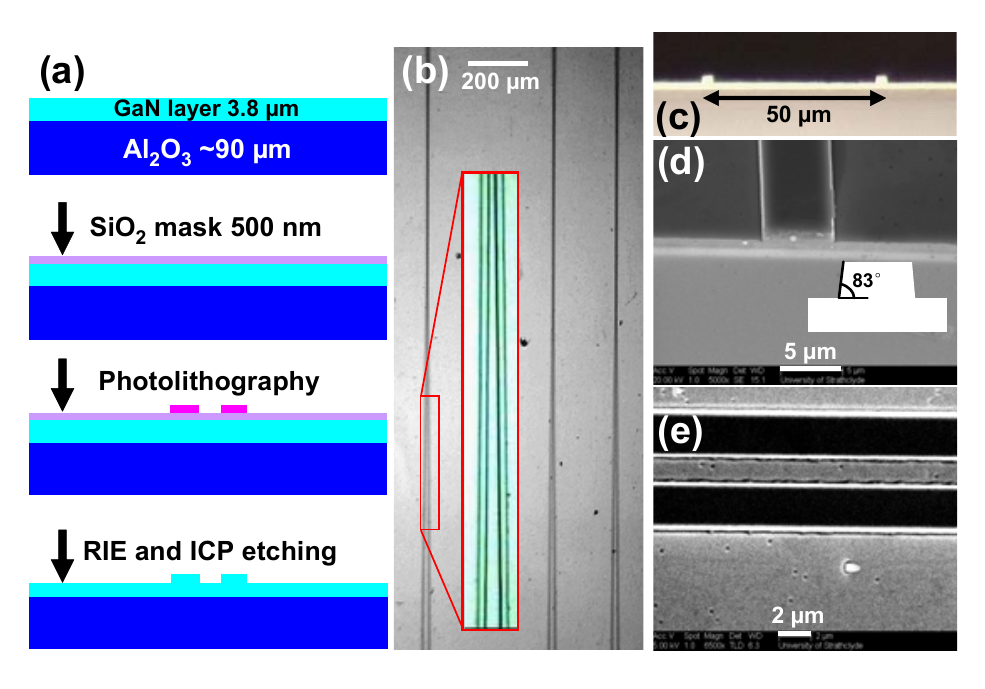}
\caption{\label{Fabrication}(Color online) GaN DC fabrication. (a) Fabrication process flow. (b) Optical plan view micrograph of GaN DCs with (inset) high magnification image of the coupling region. (c) Optical image of cleaved GaN DC facet. (d) oblique SEM of cleaved facet with (inset) inferred facet etch profile. (e) SEM plan view of coupling region. Black regions indicate the waveguides. The coupler gap and waveguide sidewalls are clearly visible.}
\end{figure}

\begin{figure}
\includegraphics[width=85mm]{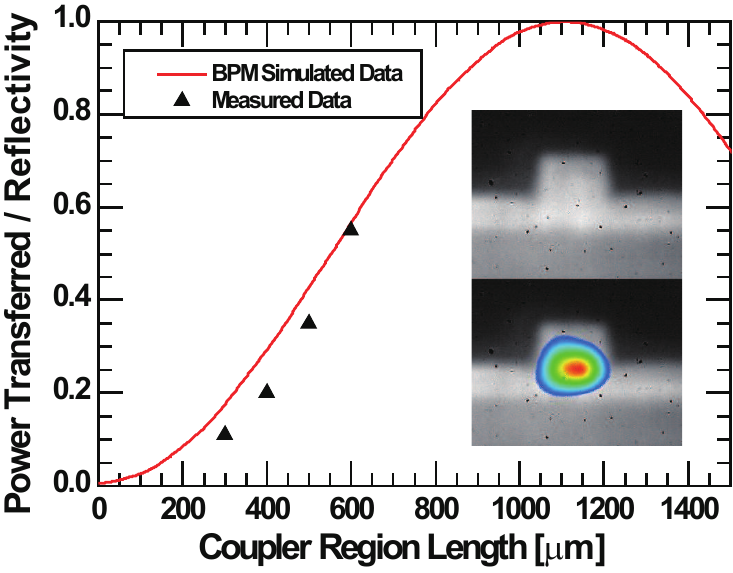}
\caption{\label{PowerTransfer}(Color online) Measured splitting ratio for fabricated GaN DCs with varying coupler length. The curve is a BPM simulation of the power transferred using the parameters shown in Fig. \ref{Simulation} at a wavelength of 800 nm. Inset: optical image of the output facet with an overlay of the measured intensity profile of the guided mode.}
\end{figure}

Device fabrication was carried out to the above specified design. The coupling section length was varied between 300 and 600 $\mu$m to obtain couplers with simulated coupling ratios ranging between 10:90 and 55:45 [Fig. \ref{PowerTransfer}]. The sapphire substrate was thinned down to 90 $\mu$m to assist cleaving, in order to provide smooth facets perpendicular to the guide for the end coupling. The DC structures were defined by optical photolithography into S1805 photoresist, transferred to a 500-nm SiO$_2$ hard mask by reactive ion etching and subsequently into GaN by inductively-coupled plasma etching using Cl$_2$/Ar gases, following the process flow of Fig. \ref{Fabrication}(a). To ease arm discrimination with free space optics the device was positioned in the center of a 5-mm-long chip with the $\sim$2 mm sections at either end of the device providing additional separation between the arms with an angular separating rate of 1.1$^o$. Stylus profilometer measurement confirms that the etching depth of the GaN coupler is 1.9 $\mu$m. This led to the structure shown in Fig. \ref{Fabrication}(b) which presents guides with various coupling lengths and (inset) a high magnification image of the coupling region. 

The sample was cleaved and optical microscopy was used to inspect the input and output facets of the GaN DCs, as shown in Fig. \ref{Fabrication}(c). SEM inspection was conducted and confirmed optical quality facets [Fig. \ref{Fabrication}(d)] and a well defined coupling region gap [Fig. \ref{Fabrication}(e)]. The sloped sidewalls were measured to be 7$^o$-off-verticality and BPM simulations were performed to confirm that this feature does not substantially change the device coupling characteristics compared to vertical sidewalls\cite{Llobera02}. It is possible to further optimize and fine tune the coupling length by accurate control of the sidewall angle with particular GaN etching recipes. 

\begin{figure}
\includegraphics[width=85mm]{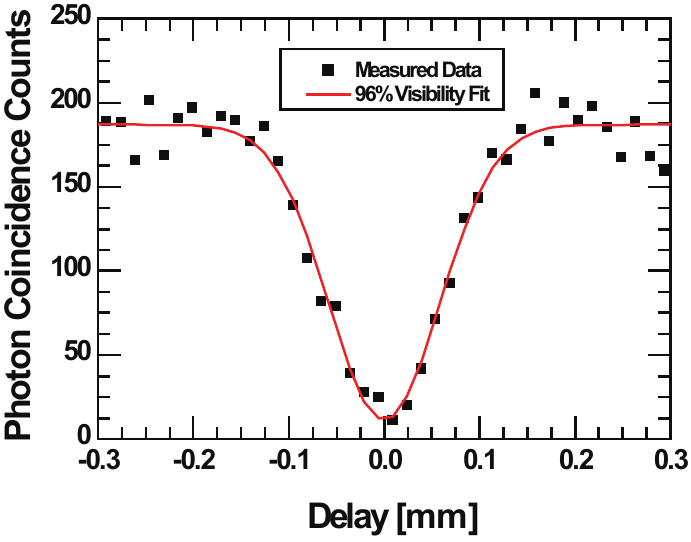}
\caption{\label{HOM}(Color online) The Hong-Ou-Mandel dip observed for a 45:55 GaN rib waveguide DC, the signature of quantum interference between two degenerate photons. The fitted visibility is 96$\%$.}
\end{figure}

The guided mode profile of the GaN straight waveguides and DCs was characterized using a diode-laser emitting at a wavelength of 800 nm. Laser light was coupled into the waveguides using free-space optics and the output from the waveguide was imaged onto a charged coupled device (CCD) with a 50X microscope objective. An image of the output facet in combination with an overlay of the measured light intensity profile is shown as inset in Fig. \ref{PowerTransfer}. This measured mode profile remained static throughout a range of coupling positions confirming the single-mode nature of the waveguide.

The splitting ratio of the DCs was measured by injecting a fixed power into each arm of the DC and measuring the power at both output  waveguide ports. This method mitigates potential loss differences between the different arms of the DC. The range of measured splitting ratios was found to vary between 10:90 to 55:45 (Fig. \ref{PowerTransfer}) and the coupler length dependence is shown to be close to simulation predictions. Using a narrow-linewidth wavelength-tunable laser source (HP 8167B) and the well established Fabry-Perot loss measurement technique\cite{Walker85}, the upper limit of propagation loss at a wavelength of 1300 nm was evaluated to be 10 dB/cm on straight waveguides fabricated by the same method. This loss can be reduced using an AlN/GaN short period-superlattice (SPS) buffer layer system during growth\cite{stolz:161903}.

Finally, to confirm the suitability of such GaN structures for integrated quantum photonic circuits, a two-photon quantum interference experiment was carried out with a DC with a 55:45 splitting ratio. As described in Ref.\onlinecite{politi08}, degenerate photon pairs at a wavelength of 804 nm were produced by spontaneous parametric down conversion using a GaN-laser-diode and a $\beta$-barium borate type-I crystal. These photon pairs were fiber-coupled to separate polarization maintaining fibers (PMF) and relayed using free-space optics to the input waveguides of the GaN DC. Output photons were collected in a similar manner and coupled to avalanche photo diodes. Photon counting and coincidence logic was then performed with field programmable gate array circuitry with computer control. Fig. \ref{HOM} shows the measured Hong-Ou-Mandel dip, the signature of quantum interference between two degenerate photon pairs\cite{Hong87}. The fitted visibility of 96$\%$ demonstrates high quality interference and confirms single-mode propagation.

In summary, GaN-on-sapphire straight and DC waveguides have been designed, fabricated and characterized with application as quantum information circuits in mind. DC design exploiting optimized waveguide cross-sections and inverse tapers for low-loss and minimal footprint structures has been proposed and experimentally validated. Single-mode devices with coupling splitting ratios varying between 10:90 and 55:45 have been studied and two-photon interference demonstrated in these strucutures. This demonstration opens the way to more sophisticated quantum photonic circuits based on GaN.

This work was supported by EPSRC, ERC, IARPA, the Leverhulme Trust, QIP IRC, QUANTIP, PHORBITECH and NSQI. J.L.O'B. acknowledges a Royal Society Wolfson Merit Award.

%

\end{document}